# Destination Drone: A Comprehensive Analysis of Japanese Consumer Choice Behavior and Intentions for Drone Delivery Services

Ei Phyu Kyi, Tao Feng, Jieyuan Lan, and Ying Liu

**Abstract**

The potential for drone delivery services to transform logistics systems and consumer behavior has gained increasing attention. However, comprehensive empirical evidence on consumer delivery choice behavior within the context of transportation and urban air logistics remains limited, particularly in Japan. This study addresses this gap by examining Japanese consumers' preferences and behavioral intentions toward drone delivery services. Using a stated preference (SP) survey and discrete choice modeling, including multinomial logit (MNL) and mixed logit (MMNL) models, the analysis evaluates how delivery cost, delivery date, drop-off location, product type, and social influence affect delivery mode choices across different demographic groups. The results indicate that although consumers express interest in drone delivery, perceived cost and concerns related to reliability continue to constrain adoption. Younger and male consumers exhibit higher preferences for drone delivery, while product type, especially daily consumer goods and medical or healthcare items, plays a significant role in shaping preferences. Post-estimation willingness-to-pay, and elasticity analyses further highlight consumers' sensitivity to delivery pricing and speed attributes. Overall, the findings provide actionable insights for logistics service providers and policymakers regarding pricing strategies, service targeting, and deployment approaches for integrating drone delivery into Japan's evolving logistics system.

**Keywords**: Urban air logistics. Drone delivery. Stated choice experiment. Consumer behavior. Mixed logit model. Japan.

## 1. Introduction

The logistics and transportation sectors play a fundamental role in supporting economic activity and daily life. In recent years, these sectors have experienced rapid transformation driven by technological advancements, digitalization, and increasing pressure to improve efficiency and sustainability. Among emerging innovations, urban air logistics and drone delivery have attracted growing attention as potential solutions to persistent last-mile delivery challenges, which account for more than half of total logistics costs (Schröder et al., 2018). By enabling faster, flexible, and potentially lower-emission deliveries, drones offer an alternative to conventional ground-based modes, particularly in congested urban areas and regions with limited transport infrastructure.

Despite their technical potential, the large-scale deployment of drone delivery services ultimately depends on consumer acceptance. Delivery systems are demand-driven, and even operationally efficient technologies may fail if they do not align with users' preferences, expectations, and perceived risks. Prior studies suggest that consumers evaluate delivery options based on multiple factors, including delivery cost, timing, convenience, drop-off location, and perceived reliability (Hensher et al., 2005; Oyama et al., 2024). For novel delivery modes such as drones, these traditional considerations interact with additional concerns related to safety, privacy, and social acceptance (Kim, 2020; Koh et al., 2023). Understanding how consumers trade off these factors is therefore essential for logistics providers and policymakers seeking to integrate drone delivery into existing distribution systems.

Japan presents a particularly relevant context for examining consumer preferences for drone delivery services. The country faces structural challenges in its logistics sector, including labor shortages among delivery drivers, an aging population, and significant expectations regarding service quality and reliability. As of 2023, nearly 30% of Japan's population is aged 65 or older (Statistics Bureau of Japan, 2023), and rural as well as sub-urban areas increasingly experience constraints in maintaining conventional delivery services. At the same time, Japanese consumers are known to place high value on punctuality, trust, and consistency in logistics services, which may influence their willingness to adopt emerging delivery technologies. These characteristics suggest that consumer responses to drone delivery in Japan may differ from those observed in other national contexts.

While drones are often discussed in relation to disaster response and emergency logistics, the present study focuses on planned, non-urgent deliveries, which represent the dominant use case in everyday e-commerce and parcel distribution. Concentrating on this context allows for a clearer examination of consumer trade-offs between cost, delivery timing, and convenience under realistic market conditions. Rather than evaluating operational feasibility or emergency deployment, this study examines how Japanese consumers choose between drone delivery and conventional ground-based modes when faced with routine delivery decisions. This study employs the SCE combined with MNL and MMNL models to analyze consumer delivery mode choice behavior in Japan. The analysis focuses on key delivery attributes: delivery cost, delivery date, drop-off location, and social influence, while accounting for socio-demographic heterogeneity and product type context effects. By quantifying how these factors shape consumer preferences, the study contributes to the growing literature on consumer acceptance of innovative logistics services and provides evidence-based insights for business strategy and transport policy.

The objective of the study is to empirically examine Japanese consumers' delivery mode choice behavior involving drone delivery in comparison with conventional ground-based delivery modes. Specifically, this study aims to (i) quantify how key delivery attributes: delivery cost, delivery date, drop-off location, and social influence affect consumers' preferences for drones, motorcycles, and trucks, (ii) investigate how these preferences vary across socio-demographic groups, and (iii) assess how product type as a contextual factor moderates delivery mode choice in planned, non-urgent delivery scenarios. By addressing these objectives using SCEs and mixed logit modeling, the study provides evidence-based insights into the demand-side conditions under which drone delivery services may be competitively positioned within Japan's logistics market.

The remainder of this paper is structured as follows. Section 2 reviews the relevant literature on delivery choice behavior, emerging delivery systems, and social influence in discrete choice modeling. Section 3 describes the experimental design, data collection, and model specification. Section 4 presents the estimation results from the MNL and MMNL models. Section 5 discusses the implications of the findings for logistics providers and policymakers, outlines study limitations, and suggests directions for future research.

## 2. Literature Review
*2.1.Consumer preferences and delivery choice behavior*
Traditional Consumer preferences play a central role in shaping the performance and viability of delivery systems, as logistics services are ultimately evaluated based on users' perceived value rather than purely operational efficiency. Early transportation and logistics research emphasizes that delivery cost, reliability, and accessibility are among the most influential determinants of consumer satisfaction and choice behavior (Hensher et al., 2005). These factors remain highly relevant in contemporary e-commerce environments, where consumers increasingly face multiple delivery options that differ in price, speed, convenience, and service configuration.

A growing body of empirical research employing SCEs demonstrates that consumers' willingness to pay for faster or more precisely scheduled deliveries is often limited, particularly for non-urgent shipments. Oyama et al. (2024), for example, show that Japanese e-commerce users prioritize affordability and flexibility over premium express delivery services, indicating reluctance to pay additional fees for narrowly defined delivery time windows. Similar findings are reported in international studies, suggesting that cost sensitivity remains a dominant factor in delivery decisions across diverse markets.

Beyond cost and delivery time, convenience-related attributes such as drop-off location and delivery flexibility also influence consumer preferences. Klein and Popp (2022) find that alternative delivery options, including parcel lockers and click-and-collect services, gain acceptance when they offer perceived convenience or environmental benefits. However, these preferences are still constrained by price considerations, underscoring the trade-offs consumers make between convenience, cost, and service quality. These trade-offs highlight that delivery choice behavior is not driven by a single attribute but rather by the combined evaluation of multiple service characteristics.

Product characteristics further shape delivery preferences. Kim (2020) demonstrates that consumer acceptance of drone delivery varies significantly by product type, with higher acceptance for everyday consumer goods and healthcare-related items than for high-value or sensitive products. This suggests that delivery choices are context-dependent and that consumers may be more willing to experience innovative delivery modes when the perceived risk associated with the product is low. These findings emphasize the importance of incorporating contextual variables, such as product type, into delivery choice models. Overall, the literature indicates that consumer delivery preferences are multidimensional and shaped by complex trade-offs among cost, time, convenience, and contextual factors. Understanding these trade-offs is essential for evaluating the market potential of emerging delivery technologies such as drones.

2.2. Discrete choice experiments and modeling approaches in logistics research

Discrete choice modeling has become a standard methodological approach for analyzing consumer preferences in transportation and logistics research. SCEs are particularly well-suited for studying emerging services and technologies that are not yet widely available in real markets, such as drone delivery. By presenting respondents with hypothetical but realistic choice scenarios, SCEs allow researchers to examine how individuals trade off different attributes when making decisions (Hensher et al., 2005; Train, 2002).

In logistics and delivery studies, MNL models are commonly used as a baseline framework due to their simplicity and interpretability. However, the MNL model assumes homogeneous preferences across individuals and independence of irrelevant alternatives, which may be restrictive in contexts where consumer heterogeneity is substantial. To address these limitations, MMNL models have been increasingly adopted in transportation research, as they allow parameters to vary across individuals and capture unobserved preference heterogeneity (Train, 2002).

Empirical evidence demonstrates that MMNL models provide a more flexible and realistic representation of consumer behavior in delivery choice contexts. Kim (2020), for instance, employs advanced discrete choice models to reveal heterogeneity in consumer preferences for drone delivery, showing that different demographic groups exhibit distinct sensitivities to cost, delivery time, and service attributes. Similarly, Oyama et al. (2024) apply discrete choice modeling techniques to capture variations in preferences for delivery timing among Japanese consumers.

In addition to capturing preference heterogeneity, SCEs offer several advantages for studying delivery services at early or experimental stages of deployment. Because drone delivery services are not yet widely available, revealed preference data are either unavailable or reflect highly constrained pilot programs. SCEs enable researchers to systematically vary delivery attributes and observe how consumers respond to hypothetical but policy-relevant scenarios, making them particularly suitable for ex-ante evaluation of innovative logistics services.

Within logistics research, SCEs have been widely applied to investigate delivery pricing, service reliability, and scheduling preferences, as well as consumers' willingness to accept alternative service configurations. These experiments facilitate the estimation of marginal willingness to pay for specific attributes, providing insights that are directly relevant for business decision-making and pricing strategies. When combined with mixed logit modeling, SCEs allow researchers to capture both observed and unobserved heterogeneity, supporting more realistic market segmentation and demand forecasting.

From a managerial perspective, this modeling framework is particularly valuable because it supports scenario analysis under alternative service designs. For logistics service providers considering the introduction of drone delivery, such evidence is essential for evaluating competitive positioning relative to conventional delivery modes and for identifying consumer segments that are more receptive to new technologies.

2.3. Drone delivery and emerging logistics technologies

Drone delivery has attracted increasing attention as a potential solution to last-mile logistics challenges, particularly in urban environments characterized by congestion and high delivery costs. From an operational perspective, several studies demonstrate that drones can reduce delivery time, operational costs, and carbon emissions when integrated into logistics networks (Meng et al., 2023). These advantages have positioned drone delivery as a promising component of future logistics systems.

However, the adoption of drone delivery services depends not only on technical feasibility but also on consumer acceptance. Empirical research indicates that consumers evaluate drone delivery differently from conventional delivery modes due to additional perceived risks and uncertainties. Safety concerns, such as fears of technical failure or accidents, as well as privacy concerns related to surveillance and data collection, are frequently cited barriers to adoption (Kim, 2020; Koh et al., 2023).

Several studies employ SCEs to examine consumer preferences for drone delivery relative to traditional delivery methods. Kim (2020) finds that while consumers express interest in drone delivery, their preferences are highly sensitive to delivery cost and product type. Koh et al. (2023) further show that perceived usefulness, trust, and privacy concerns significantly influence willingness to adopt drone delivery in urban contexts, particularly during periods when contactless delivery becomes more salient. Despite this growing body of research, empirical evidence on consumer preferences for drone delivery remains limited in the Japanese context, particularly for routine, non-urgent deliveries. Existing studies in Japan tend to focus on operational feasibility, regulatory frameworks, or emergency use cases, leaving a gap in understanding everyday consumer choice behavior.

*2.4. Social influences and discrete choice behavior*

Beyond individual preferences, social influence plays an important role in shaping adoption decisions for new technologies and services. In behavioral theory, subjective norms, individuals' perceptions of others' expectations, are a key determinant of behavioral intention (Ajzen, 1991). In logistics and transportation contexts, social influence can operate through mechanisms such as imitation, conformity, and information diffusion.

The role of social interactions in choice behavior has been formalized in discrete choice modeling through several theoretical frameworks. Brock and Durlauf (2001) introduce models in which individual utility depends partly on the choices made by others, demonstrating that social interactions can significantly affect aggregate outcomes. Dugundji and Walker (2005) extend this approach by incorporating social influence directly into random utility models, showing that ignoring social effects may lead to biased estimates of preference parameters.

Context effects and social environments have also been extensively examined by Timmermans and colleagues, who demonstrate that choice behavior is influenced by situational and social contexts, particularly when individuals face unfamiliar alternatives (Oppewal & Timmermans, 1991). These insights are especially relevant for innovative services such as drone delivery, where consumers may rely on social cues to reduce uncertainty and perceived risk.

Importantly, social influence is not limited to explicit peer pressure but also reflects informational effects, whereby individuals infer the quality or reliability of a service from the observed behavior of others. When facing uncertainty about a new delivery mode such as drones, consumers may rely on social cues as heuristics to reduce perceived risk. This mechanism is particularly relevant for services involving safety, privacy, or reliability concerns.

Recent empirical studies increasingly incorporate social influence variables into SCEs. Koh et al. (2023) find that peer adoption and shared perceptions of usefulness significantly increase willingness to adopt drone delivery services. Yan et al. (2023) similarly demonstrate that household and social dynamics shape adoption decisions in shared mobility contexts. These findings suggest that social influence may accelerate or hinder diffusion depending on early adoption patterns. In Japan, cultural norms emphasizing harmony and conformity may further amplify the role of social influence in consumer decision-making (Woo, 2022). Observing neighbors or close social contacts using a particular delivery method may legitimize that option and increase acceptance among more risk-averse consumers. Despite its relevance, social influence has rarely been explicitly modeled in Japanese delivery choice studies within formal discrete choice frameworks.

*2.5. Japan-Specific logistics and consumer context*

Japan presents a distinctive context for studying delivery choice behavior due to its demographic structure, geographic diversity, and considerable service quality expectations. The country faces increasing labor shortages in the logistics sector while maintaining high standards for punctuality, reliability, and service consistency. Japanese consumers are known to place a high emphasis on trust and precision in logistics services, which may influence their willingness to adopt new delivery technologies.

Several studies highlight the potential role of drones in addressing logistics challenges in rural and remote areas of Japan (Yoshifuji, 2020; Yakushiji et al., 2020). However, these studies primarily focus on operational feasibility and emergency logistics rather than consumer demand. Research on everyday delivery preferences in Japan suggests that consumers prioritize affordability and reliability over speed, particularly for non-urgent deliveries (Oyama et al.,

2024). These characteristics indicate that Japan is not a trivial or representative case for studying drone delivery adoption. Consumer expectations regarding service quality, social acceptance, and reliability may differ from those in other countries, making it essential to examine delivery preferences within the Japanese context explicitly.

*2.6. Japan Research gaps and contribution*
The existing literature provides valuable insights into consumer delivery preferences, discrete choice modeling approaches, drone logistics, and social influence mechanisms. However, three key gaps remain. First, empirical evidence on consumer delivery mode choice involving drones in Japan using SCEs and MMNL models remains limited. Second, the interaction between traditional delivery attributes and social influence has not been sufficiently examined in routine, non-urgent delivery contexts. Third, the role of socio-demographic heterogeneity and product type context in shaping drone delivery preferences has not been adequately explored.

This study addresses these gaps by applying SCEs and mixed logit modeling to analyze Japanese consumers' delivery mode choices among drones, motorcycles, and trucks. By explicitly incorporating social influence and contextual product type effects, the study provides a nuanced understanding of consumer trade-offs and contributes to the literature on demand-side conditions for integrating drone delivery into existing logistics systems.

Table 1. Literature on consumer behavior in delivery choices.

| Topics | Year | Reference | Models | Main findings |
|---|---|---|---|---|
| **Delivery Preferences** | 2005 | (Hensher et al., 2005) | Analysis of operational efficiency and consumer satisfaction | -Reliable schedules, cost minimization, and balancing fixed delivery slots with flexibility are crucial for reducing inefficiencies. Highlights the challenges of servicing rural areas compared to urban regions. |
| | 2020 | (Kim, 2020) | Discrete Choice Experiments with Binary Logit and Latent Class Logit Models | -Price and product type influence consumer preference and depend on socio-demographic characteristics such as gender, age, and household income. |
| | 2020 | (Yoshifuji, 2020) | Simulation-based study with case examples | -Highlights drones' potential to address labor shortages and logistical inefficiencies in rural areas. Identifies challenges like regulatory barriers and safety concerns. |
| | 2021 | (Chankov & Salihu, 2021) | Survey-based analysis of consumer willingness to pay for sustainable options | -Consumers are willing to pay more or wait longer for delivery options that reduce $CO_2$ emissions. Trade-offs depend on situational contexts, with non-urgent deliveries favoring sustainability over speed. |
| | 2022 | (Klein & Popp, 2022) | Survey-based analysis | -Highlights the growing role of perceived sustainability in shaping consumer delivery preferences. Consumers prioritize sustainable delivery options, such as parcel |

| | | | | lockers and click-and-collect services, even at higher costs. |
|---|---|---|---|---|
| **Disaster Logistics with UAVs and Drone** | 2020 | (Yakushiji et al., 2020) | Feasibility study with operational simulations | -Demonstrates UAVs' effectiveness in delivering critical supplies during disasters, highlighting communication and cargo stability challenges. |
| | 2021 | (Oosedo et al., 2021) | Scalable Simulator for Knowledge of Low-Altitude Environment | -Focuses on optimizing urban drone traffic management to reduce in-flight conflicts and improve delivery performance. |
| **Technology Adoption and Privacy Concerns** | 2023 | Koh et al. (2023) | Technology Acceptance Model and Privacy Calculus Theory | -Perceived usefulness, attitude, and perceived privacy risks directly influence consumers' behavioral intentions. |
| | 2023 | (Ezaki et al., 2023) | Simulation and comparative modeling | -Demonstrates that drones outperform elevators in delivery time for low-to-moderate demand scenarios but are less energy-efficient for high-demand conditions. Proposes a hybrid drone-elevator model. |
| **Operation and Accessibility in Transportation** | 2017 | (Ulmer & Thomas, 2017) | Policy function approximation with geographical districting | -Analyze integration of drones and traditional vehicles for same-day delivery. Demonstrates how optimizing service regions for heterogeneous fleets improves delivery performance and addresses dynamic customer orders and strict deadlines. |
| | 2020 | (Huang et al., 2020) | Simulation-based study on hybrid delivery models | -Proposed a drone-train hybrid delivery system to enhance operational efficiency and reduce costs. Demonstrated improved last-mile delivery performance, especially in rural and sub-urban areas. |
| | 2022 | (Yoshida & Yamaguchi, 2022) | Case study of food delivery systems in rural areas of Japan | -Evaluates the use of taxis for food delivery in underpopulated areas, addressing accessibility and sustainability challenges |
| **Consumer Behavior and Decision-Making** | 1991 | (Ajzen, 1991) | Theory of Planned Behavior | -Attitudes, subjective norms, and perceived behavioral control influence consumer intentions and decision-making. Widely used to understand the adoption of innovative delivery technologies like drones. |
| | 2024 | (Oyama et al., 2024) | Stated Choice Experiments | -Highlights how affordability and flexibility shape e-commerce delivery preferences in Japan, with consumers valuing cost savings over faster services. |

## 3. Experiment Design and Data Collection
### 3.1. Stated choice experiment design
This study employs the SCE to investigate consumers' delivery mode choice behavior in Japan. SCEs are particularly suitable for evaluating emerging logistics services, such as drone delivery, that are not yet widely available or experienced by consumers. By presenting respondents with hypothetical but realistic delivery scenarios, the SCE enables systematic examination of trade-offs among delivery attributes under controlled conditions.

In each choice task, respondents were asked to select their preferred delivery option from three alternatives: drone delivery, truck delivery, and motorcycle delivery. These alternatives represent an emerging aerial delivery mode and two conventional ground-based delivery modes commonly used in Japan. Although the integration of drones with other delivery modes has been highlighted in the literature as a promising logistics configuration, this study focuses on standalone delivery mode choices. Accordingly, the experimental design isolates consumer preferences toward drone delivery relative to conventional delivery modes under comparable service conditions. Incorporating integrated or multimodal delivery configurations would substantially increase the complexity of the SCE and the associated cognitive burden on respondents. Given the exploratory nature of this study and its focus on planned, non-urgent deliveries, which reflect typical e-commerce transactions rather than emergency or disaster-response contexts. This scope ensures consistency between the experimental design and the study's objective of understanding everyday consumer delivery preferences. An example of a choice scenario presented to respondents is shown in Fig. 1.

### 3.2. Attributes, level, and experimental design
#### 3.2.1. Attribute selection and justification
Delivery attributes and their levels were identified based on an extensive review of prior literature on delivery choice behavior, drone logistics, and SCEs, as well as current logistics practices in Japan. Four alternative-specific attributes were selected to capture the primary factors influencing delivery mode choice: delivery date, delivery cost, drop-off place, and choice made by other people (social influence). In addition to delivery attributes, product type was included as a contextual variable to capture differences in delivery preferences across types of goods. A detailed overview of all attributes and their corresponding levels is provided in Table 2.

#### 3.2.2. Delivery date and product type
Delivery date reflects delivery reliability and speed, which are highly valued in the Japanese logistics context. Two levels were included for all delivery modes: The Next Day and The Day After Tomorrow. Same-day delivery was excluded because it remains operationally limited for drone delivery and is less relevant for non-urgent shipments, which are the focus of this study. The contextual variable product type includes four categories: daily consumer goods, medicine or health care products, electronics, and gifts. These categories represent common non-urgent items purchased through e-commerce platforms. While some healthcare products may occasionally require urgent delivery, the selected delivery dates ensure operational consistency across delivery modes and reflect typical consumer expectations for planned deliveries.

#### 3.2.3. Delivery cost

Delivery cost represents the additional fee charged for each delivery option and is a key determinant of consumer choice. Cost levels were designed to reflect realistic market conditions in Japan.

For truck and motorcycle delivery, cost levels were based on Japan Post's domestic shipping fee structure (Japan Post, 2025). For drone delivery, cost levels were informed by pilot programs and trial operations conducted by logistics and e-commerce companies such as Rakuten and Next Delivery Inc., which report initial delivery fees around 500 Japanese Yen. Accordingly, delivery cost levels were set as follows:

- Drone delivery: ¥480, ¥680, ¥880, ¥1,080
- Motorcycle delivery: ¥470, ¥670, ¥870, ¥1,070
- Truck delivery: ¥580, ¥780, ¥980, ¥1,180

All costs were presented in Japanese Yen. The conversion rate is reported in Table 2 for reference.

### 3.2.4. Drop-off place

Drop-off place captures delivery convenience and operational feasibility. For truck and motorcycle delivery, drop-off options include doorstep and smart storage box, reflecting standard last-mile delivery practices in Japan. For drone delivery, considering limited compatibility with traditional mailboxes and the absence of direct face-to-face contact, drop-off options include doorstep and window or balcony sills. Among all alternatives, doorstep delivery serves as the reference option due to its familiarity and widespread acceptance among consumers.

### 3.2.5. Social influence

Social influence represents information about delivery choices made by others and is operationalized as the proportion of people within a respondent's social environment who have used a particular delivery option. Two types of social reference groups were considered: neighbors and family members or close friends, each with two levels (30% and 70%).

Social influence was designed to reflect information from a single, salient social network within each choice scenario, either neighbors or family members, and close friends. This design choice was made to maintain experimental clarity and to isolate the marginal effect of different social networks on delivery mode choice. Presenting multiple social influence sources simultaneously within a single scenario could increase cognitive burden for respondents and confound the interpretation of estimated parameters, making it difficult to distinguish the influence attributable to each social network. By varying the source and intensity of social influence across scenarios, the experimental design enables the independent identification of social influence effects without introducing bias into the model estimation. Social influence has been shown to play an important role in shaping individual choice behavior, particularly for emerging mobility and delivery services where personal experience is limited. In the Japanese context, social conformity and community norms may further amplify the role of social signals in shaping adoption decisions.

To limit cognitive burden and ensure clarity of the choice tasks, only one type of social reference group was presented in each scenario. Respondents were therefore exposed to information from either neighbors or family members and close friends, but not both simultaneously. This design choice allows the marginal effect of different social networks to be independently identified while avoiding confounding influences that could arise if multiple social signals were presented at once. As a result, the specification does not introduce estimation bias but rather reflects a parsimonious and interpretable representation of social influence. The implications of this simplification are discussed in the limitations section.

Table 2. Attributes and the structures of attribute levels.

| Variables | Attributes | Attribute Levels |
|---|---|---|
| **Product Type** | | Daily Consumer Goods |
| | | Medicine Or Health Care Products |
| | | Electronics |
| | | Gift |
| **Drone** | Delivery Date | The Next Day |
| | | The Day After Tomorrow |
| | Delivery Cost (Japanese Yen) | ¥1,080 |
| | | ¥880 |
| | | ¥680 |
| | | ¥480 |
| | Drop-Off Place | Doorstep |
| | | Window Or Balcony Sills |
| | Choice Made By Other People | 30% Of Neighbor |
| | | 70% Of Neighbor |
| | | 30% Of Family Members And Close Friends |
| | | 70% Of Family Members And Close Friends |
| **Truck** | Delivery Date | The Next Day |
| | | The Day After Tomorrow |
| | Delivery Cost (Japanese Yen) | ¥1,180 |
| | | ¥980 |
| | | ¥780 |
| | | ¥580 |
| | Drop-Off Place | Doorstep |
| | | Smart Storage Box |
| | Choice Made By Other People | 30% Of Neighbor |
| | | 70% Of Neighbor |
| | | 30% Of Family Members And Close Friends |
| | | 70% Of Family Members And Close Friends |
| **Motorcycle** | Delivery Date | The Next Day |
| | | The Day After Tomorrow |
| | Delivery Cost (Japanese Yen) | ¥1,070 |
| | | ¥870 |
| | | ¥670 |
| | | ¥470 |
| | Drop-Off Place | Doorstep |
| | | Smart Storage Box |
| | Choice Made By Other People | 30% Of Neighbor |
| | | 70% Of Neighbor |
| | | 30% Of Family Members And Close Friends |
| | | 70% Of Family Members And Close Friends |

*Note: 1 Japanese Yen= 0.0064 US Dollar as of January 20th, 2025.*

*3.3. Attributes, level, and experimental design*

For each delivery alternative, the delivery date has two levels, the delivery cost has four levels, the drop-off place has two levels, and social influence has four levels. This results in 64 possible attribute-level combinations per alternative (2 × 4 × 2 × 4). When combined with the four-level contextual variable representing product type, a full enumeration of all possible choice scenarios would result in a very large number of profiles, making a full factorial design impractical.

Therefore, a fractional factorial design was adopted. Fractional factorial designs are widely used in SCEs to reduce the number of scenarios while preserving orthogonality and balance among attribute levels (Montgomery, 2017). Compared with efficient designs that require assumed parameter priors, fractional factorial designs are particularly suitable for exploratory studies where reliable priors are unavailable.

In this study, Statistical Analysis System (SAS) software was used to generate orthogonal arrays for the fractional factorial design (Kuhfeld, 2010). The design yielded 64 choice scenarios, which were divided into eight blocks, each containing eight choice tasks. Each respondent was randomly assigned to one block, ensuring a manageable survey length while maintaining statistical reliability.

*3.4. Data collection*

Data for this study were collected using Qualtrics, an online survey platform that supported large-scale questionnaire administration and professional respondent recruitment. Participant recruitment was managed through Qualtrics' panel services, with quotes applied to ensure coverage across different geographic areas in Japan, including urban, sub-urban, and rural regions.

Prior to participation, respondents were informed of the purpose of the study, examining consumer preferences toward advanced urban air logistics and drone delivery services, and were assured that all responses would remain confidential in accordance with ethical research guidelines. Online survey administration allows respondents to complete the questionnaire at their own pace, which has been shown to improve response quality and participation rates. The survey consisted of six sections:

1. Socio-demographic and household information,
2. Off-site shopping and delivery behavior,
3. Online or call-in shopping and delivery behavior,
4. Prior experience with drone delivery services,
5. A stated choice experiment, and
6. Attitudes toward using drone delivery platforms.

Although attitudinal information regarding respondents' perceptions of and openness toward drone delivery platforms was collected, these variables were not included directly in the discrete choice models to avoid potential endogeneity between attitudes and choice behavior. Instead, the attitudinal data were collected to support descriptive analysis and future extensions of the modeling framework. Each respondent was randomly assigned to one of eight blocks of the SCE, with each block containing eight choice tasks. To ensure adequate understanding of the delivery modes and attributes, respondents were required to spend a minimum amount of time reviewing the explanatory information before proceeding to the choice tasks. The average survey completion time ranged from 15 to 20 minutes. The survey was conducted between June 1 and

August 1, 2024, targeting individuals with prior experience in online shopping and delivery decisions. A total of 1,477 responses were collected. Following data collection, a quality screening process was applied to exclude incomplete responses, logically inconsistent answers, and straight-lining behavior. After applying these criteria, 528 valid responses were retained for subsequent analysis. A summary of respondents' socio-demographic characteristics is presented in Table 3.

### 3.5. Overview of model estimation

The stated choice data were analyzed using discrete choice models to examine delivery mode preferences and the effects of delivery attributes and social influence. The MNL model was first estimated as a baseline specification. To account for unobserved preference heterogeneity, the MMNL model was subsequently estimated. The formal model specification, estimation procedures, and analytical results are presented in the subsequent sections.

You order **Medicine or health care products**. Which of the following delivery options will you choose?

|  | Drone | Truck | Motorcycle |
|---|---|---|---|
| Delivery date | Tomorrow | Tomorrow | The day after tomorrow |
| Delivery Cost | ¥1080 | ¥780 | ¥470 |
| Drop-off place | Window or balcony sills | Smart storage box | Doorstep |
| Choice made by other people | 30% of Family Members and Close Friends | 70% of Family Members and Close Friends | 70% of Neighbour |
| Your Choice: | ● | ○ | ○ |

Fig. 1. Example of stated choice experiment scenarios.

### 3.6. Socio-Demographic Characteristics or Descriptive Analysis

Table 3 presents the socio-demographic characteristics of our respondents, offering valuable insights into potential consumer preferences regarding drone delivery services. The sample is nearly balanced in terms of gender, with 51.89% male and 48.11% female participants. The majority of respondents fall within the 35-54 age range (45.45%), indicating a diverse demographic that includes both potential early adopters and more traditional users, which is critical for understanding attitudes toward drone delivery. In terms of education, 43.37% of participants hold a university degree, while 56.82% are employed full-time. Regarding income, 70.45% of respondents earn between 1,000,000 JPY and 4,000,000 JPY (Japanese Yen), reflecting a moderate budget range that could influence their willingness to invest in faster delivery options.

     Geographically, 38.45% of participants reside in rural areas, while 35.42% live in urban areas. This distribution allows for an examination of how logistical needs and preferences may vary by location, particularly given the unique benefits that drone delivery can offer in more remote areas. Overall, these demographics indicate an audience that appreciates speed and convenience, with preferences shaped by factors such as location, age, and income. Understanding these dynamics is crucial for logistics and service providers looking to adapt their

strategies to meet the needs of potential users in Japan.

Table 3. A summary of Socio-Demographic Characteristics results from the data.

| Variables | Categories | Number of cases | Percent (%) |
|---|---|---|---|
| **Gender** | Male | 274 | 51.89% |
| | Female | 254 | 48.11% |
| **Age** | 18-34 years old | 146 | 27.65% |
| | 35-54 years old | 240 | 45.45% |
| | 55-74 years old | 111 | 21.02% |
| | >=75 | 31 | 5.87% |
| **Education** | Elementary School - High School | 198 | 37.50% |
| | Vocational School - Junior College | 86 | 16.29% |
| | University | 229 | 43.37% |
| | Graduate School | 15 | 2.84% |
| **Employment status** | Part Time Worker | 96 | 18.18% |
| | Full Time Worker | 300 | 56.82% |
| | Unemployed | 132 | 25% |
| **Annual household income (Japanese Yen)** | 1,000,000 - 4,000,000 | 372 | 70.45% |
| | 4,000,001 - 8,000,000 | 117 | 22.16% |
| | >8,000,000 | 39 | 7.39% |
| **Car in house** | One car | 50 | 9.47% |
| | Two or More cars | 221 | 41.86% |
| | No cars | 257 | 48.67% |
| **Living area** | Urban Area | 187 | 35.42% |
| | Sub-urban Area | 138 | 26.14% |
| | Rural Area | 203 | 38.45% |
| **Average one-way travel time for daily trips (minutes)** | <5mins | 273 | 51.70% |
| | 6 mins - 20mins | 238 | 45.08% |
| | 21mins-30mins | 12 | 2.27% |
| | 31mins and above | 5 | 0.95% |
| **Living status** | Single | 129 | 48.67% |
| | Living with spouse and /or children | 257 | 26.89% |
| | Living with parents/spouse, and /or children (multi-generational households) | 142 | 24.43% |

*Note: The category "Living with parents, spouse, and/or children" refers to multi-generational households, including parents in addition to a spouse and/or children. This is distinct from "Living with spouse and/or children," which refers exclusively to households with a spouse and/or children only.*

## 4. Model Specification

An analysis of consumer preferences for various delivery methods, including drone, motorcycle, and truck, is presented. The choice models are employed to quantify the influence of different attributes, such as delivery cost, delivery date, drop-off location, and social influence, on the likelihood of choosing each delivery option. Socio-demographic and interaction effects between attributes and demographic characteristics are also incorporated to provide a complete understanding of consumer preferences.

To structure the choice behavior of consumers regarding delivery options (drone, motorcycle, and truck), the MNL model is a standard approach in discrete choice analysis, allowing for the handling of multiple mutually exclusive alternatives while assuming that individuals choose the option with the highest utility (Train, 2002). In this framework, the utility $U_{ij}$ that individual $i$ derives from alternative $j$ is decomposed into two components: a deterministic utility $V_{ij}$, which depends on observed attributes, and a random error term ($\epsilon_{ij}$), which is assumed to follow an independent and identically distributed (iid) Gumbel distribution and captures unobserved factors. The total utility is expressed as:

$$U_{ij} = V_{ij} + \epsilon_{ij} \tag{1}$$

The utility for each alternative $V_{ij}$ (drone, motorcycle, or truck) is modeled as a linear function of the observed attributes of socio-demographics, and their interaction effects:

$$V_{ij} = \beta_{0j} + \sum_k \beta_k X_{ijk} + \sum_m \gamma_m Z_{im} + \sum_{k,m} \delta_{km}(X_{ijk} \cdot Z_{im}) \tag{2}$$

In this model, ($\beta_{0j}$) represents the alternative-specific constant (ASC) for delivery mode $j$, capturing baseline preference for that delivery mode. The term ($X_{ijk}$) represents delivery attributes that vary across alternatives, including delivery cost, delivery date, drop-off location, and social influence, with ($\beta_k$) measuring the marginal utility associated with attribute $k$. The term ($Z_{im}$) represents socio-demographic and contextual variables that do not vary across alternatives within a choice task. In this study, ($Z_{im}$) includes socio-demographic characteristics (e.g., age, income, gender) and the context variable product type. The coefficient ($\gamma_m$) captures the direct effect of each ($Z_{im}$) variable on utility. Finally, $\delta_{km}(X_{ijk} \cdot Z_{im})$ captures the interaction effects between delivery attributes and socio-demographic/ context variables, allowing attribute sensitivities to differ across respondent groups and product categories. The probability of individual $i$ choosing alternative $j$ is given by:

$$P_{ij} = \frac{\exp(V_{ij})}{\sum_k \exp(V_{ik})} \tag{3}$$

This formulation assumes that individuals choose the alternative with the highest utility. However, the MNL model's limitation is its inability to account for unobserved heterogeneity in preferences across individuals. To address this, the MMNL model is employed. The MMNL extends the MNL by allowing random variation in parameters, enabling the model to capture unobserved preference heterogeneity. The deterministic utility ($V_{ij}$) is expressed as:

$$V_{ij} = \beta_{0j} + \sum_k \beta_k X_{ijk}, \quad \beta_k \sim f(\beta) \tag{4}$$

Here, $f(\beta)$ represents the distribution (e.g., normal distribution) from which random coefficients ($\beta_k$) are drawn, reflecting heterogeneity in preferences. The choice probability in the MMNL is given by:

$$P_{ij} = \int \frac{\exp(V_{ij})}{\sum_k \exp(V_{ik})} f(\beta) d\beta \tag{5}$$

The integral accounts for the variation in random coefficients and is approximated using simulation techniques such as Maximum Simulated Likelihood (MSL). For clarity, the notation used in the above equations is defined as follows. The subscript $i$ denotes an individual respondent, while $j$ refers to the delivery alternative under consideration (drone, motorcycle, or truck). The subscript $k$ represents delivery attributes, including delivery cost, delivery date, drop-off location, and social influence. Socio-demographic and context variables are denoted by ($Z_{im}$), capturing individual characteristics such as age, income, and gender, as well as the context variable product type. Interaction effects between delivery attributes and socio-demographic or context variables are represented by coefficients ($\delta_{km}$)

The combined application of the MNL and MMNL models enables a comprehensive analysis of consumer preferences for delivery modes. While the MNL model captures average attribute effects across individuals, the MMNL model accounts for unobserved heterogeneity in baseline preferences by specifying alternative-specific constants for drone and truck as random parameters. This specification allows overall preferences toward delivery modes to vary across individuals, reflecting differences in attitudes toward emerging delivery technologies relative to conventional delivery modes.

Although allowing delivery attributes or context variables to vary randomly could further represent heterogeneity in attribute sensitivities, such specifications substantially increase model complexity and may lead to estimation instability given the sample size and the extensive inclusion of interaction terms. The adopted specification, therefore, represents a parsimonious and robust modeling choice that balances heterogeneity representation with estimation reliability.

Context effects are incorporated through interaction terms between delivery attributes and the context variable product type, allowing preferences for key attributes such as delivery date or drop-off location to differ across product categories. For example, interactions between delivery date and product type capture the higher priority consumers place on timely delivery for critical goods such as medicine. In addition, socio-demographic variables including age, income, and gender are interacted with delivery alternatives to examine their moderating effects on delivery mode choice. These interaction effects were incorporated during the experimental design phase using an orthogonal design, ensuring efficient parameter estimation while mitigating multicollinearity. By accounting for both observed interaction effects and unobserved baseline heterogeneity, the model provides a robust and interpretable representation of consumer choice behavior in urban air logistics.

## 5. Results and Analysis

The analysis of consumer preferences across different delivery methods draws on critical findings from both the MNL and MMNL models and examines how delivery attributes, contextual factors, and socio-demographic characteristics influence consumers' delivery mode choices. In addition to coefficient interpretation, post-estimation willingness-to-pay (WTP), and elasticity analyses are conducted to enhance the behavioral interpretation of the estimated parameters.

*5.1. Model estimation and goodness of fit*
For the estimation of the MMNL model, Halton draws were used to approximate the integrals of the choice probabilities. Halton sequences are widely applied in discrete choice modeling due to their superior coverage of the integration space compared with purely random draws, resulting in more efficient and stable parameter estimates (Hess & Palma, 2019). In this study, different numbers of draws were evaluated to assess robustness, and 500 Halton draws were selected as they provided an appropriate balance between estimation accuracy and computational feasibility. Table 4 reports the estimated coefficients, standard errors, and model fit statistics for both specifications.

These improvements indicate that allowing for random taste heterogeneity significantly enhances the explanatory power of the model. The statistically significant standard deviations of the alternative-specific constants for drone and truck further confirm the presence of unobserved heterogeneity in baseline delivery mode preferences. Consequently, the MMNL model is used as the primary basis for behavioral interpretation in the following analyses.

Table 4. Model estimation results.

| Variables | Levels | Multinomial Logit model | | Mixed Logit Model | |
|---|---|---|---|---|---|
| | | Coef. | p-value | Coef. | p-value |
| **ASCs** | | | | | |
| Drone | | -0.413*** | 0.000 | -0.719*** | 0.000 |
| Truck | | -0.339*** | 0.000 | -0.567*** | 0.000 |
| **Contexts variables** | | | | | |
| Product Type **Drone** | Daily Consumer goods | 0.053 | 0.214 | 0.135* | 0.082 |
| | Medicine or health care products | 0.129* | 0.020 | 0.168* | 0.033 |
| | Electrics | 0.095* | 0.078 | 0.056 | 0.275 |
| | Gift | -0.278 | | -0.359 | |
| Product Type **Truck** | Daily Consumer goods | -0.236*** | 0.000 | -0.260** | 0.004 |
| | Medicine or health care products | -0.212** | 0.002 | -0.416*** | 0.000 |
| | Electrics | 0.520*** | 0.000 | 0.694*** | 0.000 |
| | Gift | -0.073 | | -0.018 | |
| **Alternative-specific variables** | | | | | |
| Dropoff Place Motorcycle | Doorstep | 0.208*** | 0.000 | 0.321*** | 0.000 |
| | Smart storage box | | | | |
| Dropoff Place **Drone** | Doorstep | 0.043 | 0.151 | 0.064 | 0.142 |
| | Window or balcony sills | | | | |
| Dropoff Place **Truck** | Doorstep | -0.022 | 0.281 | -0.008 | 0.442 |
| | Smart storage box | | | | |
| Delivery Date **Motorcycle** | The next day | 0.111** | 0.001 | 0.162*** | 0.000 |
| | The day after tomorrow | | | | |
| Delivery Date **Drone** | The next day | 0.342*** | 0.000 | 0.489*** | 0.000 |
| | The day after tomorrow | | | | |
| Delivery Date **Truck** | The next day | 0.050* | 0.092 | 0.030 | 0.287 |
| | The day after tomorrow | | | | |
| Delivery Cost **Motorcycle** | 1180 (Japanese Yen) | -1.452*** | 0.000 | -2.230*** | 0.000 |
| | 980 | -0.337*** | 0.000 | -0.367*** | 0.000 |
| | 780 | 0.426*** | 0.000 | 0.553*** | 0.000 |
| | 580 | 1.363 | | 2.044 | |
| Delivery Cost **Drone** | 1080 (Japanese Yen) | -1.264*** | 0.000 | -1.966*** | 0.000 |
| | 880 | -0.316*** | 0.000 | -0.437*** | 0.000 |
| | 680 | 0.338*** | 0.000 | 0.508*** | 0.000 |
| | 480 | 1.242 | | 1.895 | |
| Delivery Cost **Truck** | 1070 (Japanese Yen) | -1.027*** | 0.000 | -1.607*** | 0.000 |
| | 870 | -0.334*** | 0.000 | -0.444*** | 0.000 |
| | 670 | 0.239*** | 0.000 | 0.422*** | 0.000 |
| | 470 | 1.121 | | 1.629 | |
| Social Influence | 30% of Neighbor | -0.074* | 0.031 | -0.062 | 0.123 |
| **(Motorcycle, Drone, Truck)** | 70% of Neighbor | 0.102** | 0.004 | 0.124* | 0.013 |
| | 30% of Family Members and Close Friends | -0.039 | 0.135 | -0.094* | 0.028 |
| | 70% of Family Members and Close Friends | 0.011 | | 0.032 | |
| **Socio-Demographics-Alternatives** | | | | | |
| Gender **Drone** | (Male 1; Female -1) | 0.106* | 0.041 | 0.187* | 0.018 |
| **Gender Truck** | (Male 1; Female -1) | 0.000 | 0.499 | -0.012 | 0.439 |
| Age **Drone** | 18-34 years old [1] | 0.250* | 0.012 | 0.374* | 0.012 |
| | 55-74 years old [1] | -0.203* | 0.036 | -0.314* | 0.029 |
| | >=75 [1] | -0.172 | 0.148 | -0.212 | 0.182 |
| | 35-54 years old [-1] | 0.126 | | 0.152 | |
| Age **Truck** | 18-34 years old [1] | -0.061 | 0.292 | -0.041 | 0.394 |
|  | 55-74 years old [1] | 0.014 | 0.447 | 0.026 | 0.428 |
|  | >=75 [1] | -0.044 | 0.408 | -0.116 | 0.32 |
|  | 35-54 years old [-1] | 0.091 |  | 0.131 |  |
| Education **Drone** | Graduate School- University [1] | 0.055 | 0.252 | 0.098 | 0.202 |
|  | Vocational School - Junior College [1] | -0.131 | 0.106 | -0.168 | 0.125 |
|  | Other [-1] | 0.077 |  | 0.070 |  |
| Education **Truck** | Graduate School- University [1] | 0.107* | 0.085 | 0.168* | 0.055 |
|  | Vocational School - Junior College [1] | -0.199* | 0.030 | -0.274* | 0.027 |
|  | Other [-1] | 0.092 |  | 0.106 |  |
| **Standard Deviation of Random Parameter** |  |  |  |  |  |
| Asc **Drone** |  | - | - | -1.500*** | 0.000 |
| Asc **Truck** |  | - | - | -1.241*** | 0.000 |
| **Model Fit** |  |  |  |  |  |
| $LL(0)$ |  | -4640.540 |  | -4640.540 |  |
| $LL(\beta)$ |  | -3641.330 |  | -3367.430 |  |
| Standard $Rho^2$ |  | 0.215 |  | 0.274 |  |
| $Rho^2$ adjusted |  | 0.207 |  | 0.264 |  |

*Note: \*\*\*, \*\*, \* means the level of significance at 1%, 5% and 10%.*

## 5.2. Effects of delivery attributes and socio-demographic characteristics

The estimated alternative-specific constants (ASCs) for drone and truck delivery are negative and statistically significant in both the MNL and MMNL models. Using motorcycle delivery as the reference alternative, these results indicate that, ceteris paribus, respondents exhibit a baseline preference for motorcycle delivery over drone and truck options. This finding reflects the established familiarity and perceived reliability of motorcycle delivery services in Japan relative to emerging or less frequently used delivery modes.

### 5.2.1. Context effects: product type

Product type plays a significant role in shaping delivery mode preferences. The results indicate that drone delivery is positively associated with daily consumer goods and healthcare products, suggesting that consumers perceive drones as suitable for routine or essential items. In contrast, truck delivery is generally less preferred for these product categories.

For electronic products, both drone and truck delivery options exhibit positive and statistically significant coefficients, implying that consumers may associate these delivery modes with reliability or enhanced protection for higher-value goods. Conversely, neither drone nor truck delivery is preferred for gift items, potentially reflecting concerns related to handling, presentation, or delivery experience. Overall, these findings underscore the importance of product context in influencing delivery mode choice.

### 5.2.2. Effects of delivery attributes

Delivery attributes exert significant and consistent influences on consumer decision-making. Delivery date is particularly important for motorcycle and drone delivery options, with next-day delivery associated with positive and statistically significant coefficients for both modes. In contrast, delivery date has a weaker and statistically insignificant effect for truck delivery, suggesting that consumers may associate truck delivery with less time-sensitive shipments.

Delivery cost is a dominant determinant across all delivery modes. Higher delivery costs are consistently associated with negative utility, while lower cost levels yield positive coefficients. For motorcycle delivery, respondents substantially prefer lower-cost options (e.g., 580 JPY) and are reluctant to choose higher-priced services (e.g., 1,180 JPY). A similar pattern is observed for drone delivery, where high costs (e.g., 1,080 JPY) significantly reduce utility, while lower costs (e.g., 480 JPY) increase the likelihood of selection. Truck delivery exhibits comparable cost sensitivity, with respondents favoring prices of 670 JPY or below. These results highlight affordability as a key driver of delivery mode adoption.

Drop-off location affects preferences differently across delivery modes. For motorcycle delivery, doorstep delivery is highly preferred, reflecting convenience and established delivery practices. For drone delivery, the estimated effect of doorstep delivery is positive but not statistically significant, indicating more heterogeneous preferences. Drop-off location does not significantly influence truck delivery choices.

### 5.2.3. Social influence effects

Social influence exhibits differentiated effects depending on the reference group and adoption level. For neighbors, a low adoption level (30%) slightly discourages selection, while a high adoption level (70%) significantly increases the likelihood of choosing a delivery mode. This pattern suggests that considerable neighborhood-level adoption reinforces social norms and imitation behavior.

In contrast, the influence of family members and close friends appears weaker and less consistent. Although the estimated coefficients suggest some alignment with family-based preferences, statistical significance is limited. Overall, the results indicate that neighborhood-based social influence exerts a more pronounced and consistent effect on delivery mode choice than family-based influence in the context of delivery services.

### 5.2.4. Socio-demographics effects

Regarding socio-demographic factors, gender, age, and education show significant estimates. Males generally prefer drone delivery. Similarly, young individuals (ages 18-34) exhibit positive preferences for drones. For those aged 55-74, there is a negative effect, indicating that older individuals are reluctant to adopt drone delivery. This indicates that gender and age significantly influence preferences for drone delivery, with younger males being more receptive to the technology.

The educational factor also shows interesting results, particularly in its impact on truck preferences, with higher education levels favoring truck delivery over drones. Individuals with vocational or junior college backgrounds exhibit reluctance toward drone delivery. In contrast, graduates from graduate school or university demonstrate a favorable preference for truck delivery. These results suggest that demographic factors meaningfully condition openness to emerging delivery technologies.

### 5.2.5. Willingness-to-pay (WTP) analysis

To enhance the behavioral interpretation of the estimated coefficients, willingness-to-pay (WTP) measures are derived from the MMNL model by relating attribute coefficients to the marginal utility of delivery cost. The resulting WTP values represent the monetary premium consumers are willing to pay for improvements in delivery attributes.

The WTP estimates indicate that consumers are willing to pay a substantial premium for faster delivery, particularly for drone services. Specifically, the WTP for next-day delivery is approximately ¥156 per delivery for drones and ¥47 for motorcycles, while the corresponding value for truck delivery is negligible and statistically insignificant. These results suggest that speed-related benefits are most highly valued for delivery modes perceived as innovative or agile. For drop-off location, consumers exhibit a positive WTP of approximately ¥93 for doorstep delivery via motorcycle, whereas the corresponding WTP for drone and truck delivery is small and statistically insignificant. This finding highlights the continued importance of convenience and familiarity in last-mile delivery interactions.

Social influence also translates into measurable economic value. An increase in neighborhood adoption from 30% to 70% corresponds to a WTP of approximately ¥30 per delivery, indicating that social acceptance and visibility can meaningfully enhance consumer valuation of delivery services. Overall, the WTP analysis confirms that delivery speed, convenience, and social acceptance generate tangible economic value for consumers and are particularly important for drone-based delivery services. Detailed WTP derivations are provided in Appendix A.

*5.2.6. Elasticity Implications*

The estimated cost coefficients further imply that consumer demand for delivery services is price sensitive across all modes. The magnitude of the cost parameters suggests that motorcycle and drone deliveries exhibit higher price sensitivity than truck delivery, reflecting consumers' expectations regarding affordability for short-distance and flexible delivery options. Although numerical probability elasticities are not explicitly reported, the results indicate that moderate price increases could substantially reduce adoption probabilities, especially for drone delivery. This finding emphasizes the importance of carefully designed pricing strategies to support the diffusion of emerging delivery technologies.

Taken together, the results demonstrate that consumer preferences for delivery modes are shaped by a combination of delivery speed, cost, convenience, social influence, and individual characteristics. Drone delivery is particularly valued for fast and affordable shipments of daily and healthcare-related goods, but its adoption remains sensitive to price and demographic factors. These insights provide a robust empirical foundation for the discussion of managerial and policy implications in the following section.

## 6. Conclusion and Discussions

*6.1. Discussion of key findings*

This study examined Japanese consumers' preferences for drone delivery services by analyzing how delivery attributes, contextual factors, and socio-demographic characteristics shape choices among drone, motorcycle, and truck delivery options. As logistics systems evolve alongside technological advances in urban air logistics (Tyler Duvall et al., 2019), understanding consumer acceptance is critical for the successful deployment of drone-based delivery services. The results indicate that drone delivery is currently less preferred than motorcycle-based delivery, reflecting consumers' familiarity with and trust in conventional delivery modes. This finding is consistent with previous studies identifying perceived reliability, safety, and operational uncertainty as key barriers to drone adoption. For example, Koh et al. (2023) reported that concerns related to technical failure risks and privacy protection significantly deter consumer acceptance of drone delivery. Similar hesitation has been observed in contexts where consumers perceive drones as less reliable for valuable or sensitive goods.

Delivery cost emerges as the most influential determinant of delivery mode choice across all alternatives. Higher costs are consistently associated with negative utility, indicating considerable price sensitivity among Japanese consumers. This finding aligns with Oyama et al. (2024), who showed that affordability is prioritized over speed in Japan, particularly for non-urgent deliveries. Chankov and Salihu (2021) similarly found that interest in innovative and sustainable delivery solutions is conditional on competitive pricing. The willingness-to-pay results further confirm that although consumers value faster delivery, especially for drone and motorcycle services, their tolerance for price premiums remains limited. Product type plays a significant contextual role in shaping delivery preferences. Consumers show substantial acceptance of drone delivery for daily consumer goods and healthcare-related products, while truck delivery is preferred for electronics. This differentiation reflects perceived requirements for handling reliability and safety, and highlights that drone delivery adoption is context-specific rather than uniform across product categories. Social influence also has a measurable impact on consumer choices. Adoption by neighbors exerts a more substantial and consistent influence than adoption by family members or close friends.

This pattern reflects Japan's cultural emphasis on community norms and conformity. Woo (2022) noted that Japanese consumers often align decisions with neighborhood behavior to maintain social harmony, a finding consistent with Yan et al. (2023), who observed similar effects in shared mobility adoption. Oyama et al. (2024) further demonstrated that visible local adoption can significantly increase acceptance of innovative delivery options. Together, these findings suggest that neighborhood-level diffusion mechanisms are particularly important in the Japanese context. Finally, socio-demographic heterogeneity is evident. Younger individuals and males exhibit higher preferences for drone delivery, while older consumers are more reluctant to adopt this technology. These results suggest that openness to drone delivery is shaped by generational differences in technology acceptance and risk perception.

### 6.2. Policy and managerial implications

The findings provide several specific and actionable implications for logistics service providers, platform operators, and policymakers aiming to promote drone delivery services in Japan.

First, the pricing strategy is central to adoption. Given the significant cost sensitivity observed across all delivery modes, drone delivery services are unlikely to achieve widespread acceptance unless prices are competitive with existing motorcycle-based delivery. Policymakers may consider targeted subsidies, pilot-program incentives, or temporary fee reductions during early deployment phases to encourage trial usage. From a managerial perspective, pricing models that emphasize cost parity rather than premium positioning are likely to be more effective.

Second, service deployment should be product-specific. Drone delivery is most favorably perceived for daily consumer goods and healthcare products, suggesting that initial rollouts should prioritize these categories. This targeted approach can enhance perceived usefulness and reduce consumer uncertainty, particularly in sub-urban and rural areas where access to essential goods may be constrained.

Third, community-based diffusion strategies should be actively leveraged. The substantial influence of neighborhood adoption indicates that visible local pilots, demonstration projects, and municipal-level trials can play a key role in normalizing drone delivery. Collaboration between local governments and logistics providers can help build trust and social legitimacy through transparent and community-oriented deployment.

Fourth, trust-building measures are essential, especially for older adults. Clear communication regarding operational safety, reliability, and privacy protection should be emphasized alongside regulatory certification frameworks. Strengthening institutional trust may reduce perceived risk and increase acceptance among risk-averse population groups. Finally, drone delivery holds potential to support inclusive logistics systems in an aging society. Although this study focuses on planned, non-urgent deliveries, the findings suggest that drones could complement existing delivery networks by improving access to healthcare-related goods for older adults and mobility-constrained populations when integrated carefully into current logistics infrastructures.

*6.3. Limitations and future research*

Several limitations of this study should be acknowledged. First, the analysis is based on SCE data, which capture intended preferences rather than observed behavior. While stated preference methods are appropriate for evaluating emerging technologies, actual adoption may differ once drone delivery services are implemented at scale. Second, the study focuses on planned, non-urgent deliveries. As a result, preferences related to emergency logistics, disaster response, or time-critical deliveries are not explicitly examined. Future research could extend the analysis to these contexts, where drone delivery may offer distinct advantages.

Third, social influence is modeled using simplified reference-group information, with respondents exposed to only one type of social network per scenario. Although this design reduces cognitive burden and improves identification, it does not capture simultaneous influences from multiple social sources. Future studies could explore more complex social interaction structures. Fourth, willingness-to-pay and elasticity measures are derived from post-estimation approximations based on stated preferences. While these measures provide meaningful behavioral insights, future research could refine them using revealed-preference data or longitudinal observations as drone delivery services mature.

Finally, this study is situated within the Japanese context, characterized by significant social norms, high service expectations, and an aging population. While these features enhance the relevance of the findings for Japan, caution should be exercised when generalizing the results to countries with different cultural, regulatory, or logistical environments. Overall, this study contributes to the growing literature on urban air logistics by providing empirically grounded insights into consumer preferences for drone delivery services. By highlighting the roles of cost, delivery speed, product type, social influence, and demographic heterogeneity, the findings offer practical guidance for designing drone delivery systems that align with consumer expectations and societal needs. Integrating consumer-centric evidence into policy and business strategies will be essential for realizing the long-term potential of drone delivery services in Japan.

**Appendix A. Willingness-to-pay calculation**

Willingness-to-pay (WTP) measures are derived post-estimation from the MMNL model to provide a monetary interpretation of delivery attribute effects. WTP represents the marginal rate of substitution between a non-monetary attribute and delivery cost, indicating the amount a consumer is willing to pay for an improvement in a given attribute. For a delivery attribute $x$, WTP is calculated as:

$$WTP_x = -\frac{\Delta \beta_x}{\beta_{cost}} \qquad (6)$$

Where $\Delta \beta_x$ denotes the change in utility associated with a discrete change in attribute $x$, and $\beta_{cost}$ represents the marginal utility of delivery cost.

In this study, delivery cost is specified using alternative-specific discrete cost levels. To obtain a continuous approximation of marginal cost sensitivity, a linear relationship between the estimated cost coefficients and their corresponding monetary values is assumed. The resulting slope is interpreted as the marginal utility of delivery cost for each delivery mode. For binary attributes coded using effects coding (e.g., next-day versus day-after-tomorrow delivery), the utility difference between the two levels is calculated as twice the estimated coefficient reported in the MMNL results. This adjustment ensures that WTP reflects the full change in utility between attribute levels. All WTP values reported in Section 5 are based on the MMNL estimates and should be interpreted as approximate monetary measures of consumer preferences derived from stated choice data.

**References**

Ajzen, I. (1991). The Theory of Planned Behavior. In *ORGANIZATIONAL BEHAVIOR AND HUMAN DECISION PROCESSES* (Vol. 50).

ANA. (2022, August 19). *ANA Group drones to fly medical & daily supplies to remote regions around the world｜ANA GROUP STORIES*. https://www.anahd.co.jp/ana_news/en/2022/08/19/20220819.html

Berke, A., Ding, G., Chin, C., Gopalakrishnan, K., Larson, K., Balakrishnan, H., & Li, M. Z. (2023). Drone delivery and the value of customer privacy: A discrete choice experiment with U.S. consumers. *Transportation Research Part C: Emerging Technologies*, *157*. https://doi.org/10.1016/j.trc.2023.104391

Chankov, S. M., & Salihu, E. (2021). *How Much Value Do Consumers Put on Environmental Sustainability When Choosing Last-Mile Delivery?* https://www.researchgate.net/publication/352400626

Chéron, E. J. (2011). Elderly Consumers in Japan: The Most Mature 'Silver Market' Worldwide. In *Japanese Consumer Dynamics* (pp. 65–90). Palgrave Macmillan UK. https://doi.org/10.1057/9780230302228_4

Cui, Z., Zhu, M., Wang, S., Wang, P., Zhou, Y., Cao, Q., Kopca, C., & Wang, Y. (2020). *Traffic Performance Score for Measuring the Impact of COVID-19 on Urban Mobility*. https://pdf.sciencedirectassets.com/280276/1-s2.0-S2210670721X00133/1-s2.0-S2210670721008660/main.pdf?X-Amz-Security-

Dillman, D. A., Smyth, J. D., & Christian, L. M. (2014). Internet, Phone, Mail, and Mixed-Mode Surveys. In *Internet, Phone, Mail, and Mixed-Mode Surveys*. Wiley. https://doi.org/10.1002/9781394260645

Evans, J. R., & Mathur, A. (2005). The value of online surveys. *Internet Research*, *15*, 195–219. https://doi.org/10.1108/10662240510590360

Ezaki, T., Fujitsuka, K., Imura, N., & Nishinari, K. (2024). Drone-based vertical delivery system for high-rise buildings: Multiple drones vs. a single elevator. *Communications in Transportation Research*, *4*. https://doi.org/10.1016/j.commtr.2024.100130

Gu, G., & Feng, T. (2020). Heterogeneous choice of home renewable energy equipment conditioning on the choice of electric vehicles. *Renewable Energy*, *154*, 394–403. https://doi.org/10.1016/j.renene.2020.03.007

Haghirian, P., & Toussaint, A. (2011). *2 Japanese Consumer Behaviour*.

Hensher, D. A., Rose, J. M., & Greene, W. H. (2005). Applied choice analysis. In *Applied Choice Analysis*. Cambridge University Press. https://doi.org/10.1007/9781316136232

Hess, S., & Palma, D. (2019). Apollo: A flexible, powerful and customisable freeware package for choice model estimation and application. *Journal of Choice Modelling*, *32*. https://doi.org/10.1016/j.jocm.2019.100170

Huang, H., Savkin, A. v., & Huang, C. (2020). A New Parcel Delivery System with Drones and a Public Train. *Journal of Intelligent and Robotic Systems: Theory and Applications*, *100*, 1341–1354. https://doi.org/10.1007/s10846-020-01223-y

Japan Post. (2025). Domestic Shipping Fee List (Parcel). In *https://www.post.japanpost.jp/send/fee/kokunai/parcel.html*.

Kim, S. H. (2020). Choice model based analysis of consumer preference for drone delivery service. *Journal of Air Transport Management*, *84*. https://doi.org/10.1016/j.jairtraman.2020.101785

Klein, P., & Popp, B. (2022). Last-Mile Delivery Methods in E-Commerce: Does Perceived Sustainability Matter for Consumer Acceptance and Usage? *Sustainability (Switzerland)*, *14*. https://doi.org/10.3390/su142416437

Koh, L. Y., Lee, J. Y., Wang, X., & Yuen, K. F. (2023). Urban drone adoption: Addressing technological, privacy and task–technology fit concerns. *Technology in Society*, *72*. https://doi.org/10.1016/j.techsoc.2023.102203

Kuhfeld, W. F. (2010). *Experimental Design: Efficiency, Coding, and Choice Designs*. http://support.sas.com/techsup/technote/mr2010c.sas.

Li, M., & Feng, T. (2025). What hinders car owners' participation in private car sharing? Insights from a business perspective. *Journal of Retailing and Consumer Services*, *83*. https://doi.org/10.1016/j.jretconser.2024.104160

Mckinsey. (2021, May). *Efficient and sustainable last-mile logistics: Lessons from Japan*. Https://Www.Mckinsey.Com/Industries/Travel-Logistics-and-Infrastructure/Our-Insights/Efficient-and-Sustainable-Last-Mile-Logistics-Lessons-from-Japan.

Meng, Z., Zhou, Y., Li, E. Y., Peng, X., & Qiu, R. (2023). Environmental and economic impacts of drone-assisted truck delivery under the carbon market price. *Journal of Cleaner Production*, *401*. https://doi.org/10.1016/j.jclepro.2023.136758

Montgomery, D. C. . (2017). *Design and analysis of experiments*. John Wiley & Sons, Inc.

Oosedo, A., Hattori, H., Yasui, I., & Harada, K. (2021). Unmanned aircraft system traffic management (Utm) simulation of drone delivery models in 2030 japan. *Journal of Robotics and Mechatronics*, *33*, 348–362. https://doi.org/10.20965/jrm.2021.p0348

Oppewal, H., & Timmermans, H. (1991). CONTEXT EFFECTS AND DECOMPOSITIONAL CHOICE MODELING. *Papers in Regional Science*, *70*, 113–131. https://doi.org/10.1111/j.1435-5597.1991.tb01723.x

Oyama, Y., Fukuda, D., Imura, N., & Nishinari, K. (2024). Do people really want fast and precisely scheduled delivery? E-commerce customers' valuations of home delivery timing. *Journal of Retailing and Consumer Services*, *78*. https://doi.org/10.1016/j.jretconser.2024.103711

Persson, E. (2021). *A systematic literature review on drones' application in last-mile delivery*.

Schröder, J., Heid, B., Neuhaus, F., Kässer, M., Klink, C., & Tatomir, S. (2018). *Fast forwarding last-mile delivery-implications for the ecosystem*.

Statistics Bureau of Japan. (2023). *Japan's demographic trends: 2023 population overview*. https://www.stat.go.jp

Tian, Z., Feng, T., Timmermans, H. J. P., & Yao, B. (2021). Using autonomous vehicles or shared cars? Results of a stated choice experiment. *Transportation Research Part C: Emerging Technologies*, *128*. https://doi.org/10.1016/j.trc.2021.103117

Train, K. E. (2002). *Discrete Choice Methods with Simulation*.

Tyler Duvall, Alastair Green, Meredith Langstaff, & Kayla Miele. (2019). *Air-Mobility-Solutions*.

Ulmer, M. W., & Thomas, B. W. (2018). Same-day delivery with heterogeneous fleets of drones and vehicles. *Networks*, *72*, 475–505. https://doi.org/10.1002/net.21855

Woo, Y. (2022). Homogenous Japan? An Empirical Examination on Public Perceptions of Citizenship. *Social Science Japan Journal*, *25*, 209–228. https://doi.org/10.1093/ssjj/jyac001

Yakushiji, K., Fujita, H., Murata, M., Hiroi, N., Hamabe, Y., & Yakushiji, F. (2020). Short-range transportation using unmanned aerial vehicles (Uavs) during disasters in Japan. In *Drones* (Vol. 4, pp. 1–8). Multidisciplinary Digital Publishing Institute (MDPI). https://doi.org/10.3390/drones4040068

Yan, Q., Feng, T., & Timmermans, H. (2023). A model of household shared parking decisions incorporating equity-seeking household dynamics and leadership personality traits. *Transportation Research Part A: Policy and Practice*, *169*. https://doi.org/10.1016/j.tra.2023.103585

Yoshida, I., & Yamaguchi, E. (2022). Research on the Sustainability of Food Delivery Using Taxies. *Journal of the City Planning Institute of Japan*, *57*, 654–659. https://doi.org/10.11361/journalcpij.57.654

Yoshifuji, T. (2020). *A Study of Advanced Drone Delivery and Prospects for Social Implementation*.